# Correlated Optical Convolutional Neural Network with "Quantum Speedup"

Yifan Sun, Qian Li, Ling-Jun Kong, and Xiangdong Zhang*

*Key Laboratory of advanced optoelectronic quantum architecture and measurements of Ministry of Education, Beijing Key Laboratory of Nanophotonics & Ultrafine Optoelectronic Systems, School of Physics, Beijing Institute of Technology, 100081 Beijing, China.*

*\*Author to whom any correspondence should be addressed: zhangxd@bit.edu.cn*

## Abstract

Compared with electrical neural networks, optical neural networks (ONNs) have the potentials to break the limit of the bandwidth and reduce the consumption of energy, and therefore draw much attention in recent years. By far, several types of ONNs have been implemented. However, the current ONNs cannot realize the acceleration as powerful as that indicated by the models like quantum neural networks. How to construct and realize an ONN with the quantum speedup is a huge challenge. Here, we propose theoretically and demonstrate experimentally a new type of optical convolutional neural network by introducing the optical correlation. It is called the correlated optical convolutional neural network (COCNN). We show that the COCNN can exhibit "quantum speedup" in the training process. The character is verified from the two aspects. One is the direct illustration of the faster convergence by comparing the loss function curves of the COCNN with that of the traditional convolutional neural network (CNN). Such a result is compatible with the training performance of the recently proposed quantum convolutional neural network (QCNN). The other is the demonstration of the COCNN's capability to perform the QCNN phase recognition circuit, validating the connection between the COCNN and the QCNN. Furthermore, we take the COCNN analog to the 3-qubit QCNN phase recognition circuit as an example and perform an experiment to show the soundness and the feasibility of it. The results perfectly match the theoretical calculations. Our proposal opens up a new avenue for realizing the ONNs with the quantum speedup, which will benefit the information processing in the era of big data.

## Introduction

Artificial neural networks are the computational models composed of interconnected nodes, and can 'learn' to deal with complicated tasks such as image feature recognition, language translation, medical diagnosis, etc., through 'training' the parameters[1–4]. The optical neural networks (ONNs) can perform the function of the artificial neural networks by using optical elements. They have drawn much attentions in recent years because of the potential to go beyond their electrical counterparts. The advantages of the ONNs include the low crosstalk, neglectable time-delay in propagation, low heat generation, etc.[5,6]. Especially, they are expected to break the bandwidth limits of the electrical neural networks, achieving ultrahigh computing frequency enabled by THz-wide telecommunications band[7]. Also, the ONNs can get rid of the troubles caused by the Von Neumann bottleneck, avoiding the restrictions rooted in the energy and time consumption when reading and transmitting data from the memories[8]. With these benefits, the ONNs are proven to perform the image processing[9-16], hand-written digits recognition[17-21], and many other tasks[22-26]. The related techniques are also found to be applicable for logic[27] or matrix[28-30] operations. However, all the ONNs at present cannot exhibit a speedup as powerful as that indicated by the models such as quantum neural networks[31-39], because they are implemented in a straightforward manner that closely follows the traditional neural networks without incorporating an algorithmic advances. With the proliferation of the data generated by the daily communication, the traditional neural networks must contain millions of parameters in order to capture the feature of the data, so that the training of them for practical use is becoming more and more consumptive. Hence, the same obstacle will also occur in the current development direction of the ONNs. So, how to build a new ONN which could provide an algorithmic speed up is highly demanded by the social needs and generally challenging.

On the other hand, the recent experimental progress towards the realization of quantum information processors has led to the emerged research field of quantum neural networks[31]. Excitingly, the advance of the quantum machine learning would fulfill the needs for data processing with an algorithmic speed up. By far, several quantum machine learning algorithms are proposed, including the quantum generative network[34], quantum Boltzmann machine[35], quantum transfer learning[38], etc. Very recently, the quantum convolutional neural network (QCNN) is theoretically constructed[37]. Such a network has displayed its unique property for identifying the symmetry of quantum states. More importantly, it has been numerically shown that the convergence of a QCNN model is faster than the traditional convolutional neural network (CNN) model in the task of classifying the classical data[39], which would greatly contribute to the application in daily life. However, the realization of the QCNN requires large amount of the 2-qubit gates and the sufficient long coherence time of the multi-qubit system, which is technically hard for the current quantum devices. So, the demonstration of the advantageous quantum machine learning models like QCNN on the quantum platforms has not been given.

Inspired by the theoretical model of the QCNN, in this work, we introduce the optical correlation to the design of ONNs and propose the correlated optical convolutional neural networks (COCNNs). Such a new type of ONNs exhibits the ability to achieve the "quantum speed up" as effective as that shown by the QCNN. Such a speed up is verified from two aspects. For the first, the lost function of the COCNN has shown a faster convergence behavior in the classification tasks we consider when compared to that of the traditional CNN model. The results coincides with behavior of the QCNN[39]. For the second, the COCNN can realize the function of the QCNN, such as the recognition of the Haldane phase. This further validates the close relation between the COCNN and the QCNN. Last but not least, we take the 3-qubit phase recognition circuit as an example and perform an experimental realization of its function using the framework of the COCNN. The results fit well with the theoretical calculation. It means that our proposal reveals a new way of realizing the ONNs with the "quantum speedup", which will benefit the information processing in the coming years.

## Results

### The structure of a COCNN and its performances

The sketch of a general COCNN is shown in Fig. 1. It is composed of four basic parts: the correlated light source, the convolution, the pooling, and the detections. As the most basic component, we firstly introduce the part of the correlated light source. This setup is used for encoding the information. Different from the previous ONNs[9,13,15] which only employ the amplitude of the light for the encoding, we employ the correlation of the multi-mode polarized beams to encode data. In fact, using such a special type of classical beams, one can obtain the classical analogy of the quantum correlations between qubits. For example, a polarized beam field $\boldsymbol{E} = \alpha \boldsymbol{h} + \beta \boldsymbol{v}$, with $\boldsymbol{h}$ ($\boldsymbol{v}$) being the horizontal (vertical) polarization vector, is analog to the qubit state $\alpha|0\rangle + \beta|1\rangle$ under the mapping $\boldsymbol{h} \to |0\rangle$ and $\boldsymbol{v} \to |1\rangle$. Here, $\alpha$ and $\beta$ represent the complex coefficients of the polarization basis. In what follows, we use the notation $|\ )$ to denote the classical states analog to the qubit states, e.g., $|\boldsymbol{E}) = \alpha|\boldsymbol{h}) + \beta|\boldsymbol{v})$. More generally, by employing $N$ multi-mode polarized beams, one can obtain a classical state

$$|NE) = c_{h_1 h_2 \ldots h_N} |\boldsymbol{h}_1)|\boldsymbol{h}_2) \ldots |\boldsymbol{h}_N) + c_{h_1 h_2 \ldots v_N} |\boldsymbol{h}_1)|\boldsymbol{h}_2) \ldots |\boldsymbol{v}_N) + \cdots + c_{v_1 v_2 \ldots v_N} |\boldsymbol{v}_1)|\boldsymbol{v}_2) \ldots |\boldsymbol{v}_N) \quad (1)$$

which is the analogy of a general $N$-qubit quantum state $|\psi_N\rangle = q_{00\ldots0} |00\ldots0\rangle + q_{00\ldots1} |00\ldots1\rangle + \cdots + q_{11\ldots1} |11\ldots1\rangle$. The $N$ multi-mode polarized beams that give the state of Eq. (1) can be denoted by $\boldsymbol{E}_i$ ($i = 1, \ldots, N$), each of which can be expressed by $\sum_{k=1}^{M} f_k \boldsymbol{p}_{i,k}$. The $f_k$ of $\boldsymbol{E}_i$ denotes the orthonormal modes satisfying the relation $\int f_{k_1}(\boldsymbol{r},t) \ldots f_{k_M}(\boldsymbol{r},t) d\Omega = \delta_{k_1,\ldots,k_M}$, and $\Omega$ represents the parameter domain where the condition holds. $\boldsymbol{p}_{i,k} = p_{i,k}^{\mathrm{H}} \boldsymbol{h} + p_{i,k}^{\mathrm{V}} \boldsymbol{v}$ denotes the polarization of the mode $f_k$. Such a setup of the beams $\boldsymbol{E}_i$ ($i = 1, \ldots, N$) encodes the information to be processed, which is schematically shown in the leftmost part of Fig. 1. Fundamentally, the $N$ multi-mode polarized beams can be used to analogize the $N$-qubit state, because the correlation of beams defined by the integral $\int (\boldsymbol{e}_1 \cdot \boldsymbol{E}_1)(\boldsymbol{e}_2 \cdot \boldsymbol{E}_2) \ldots (\boldsymbol{e}_N \cdot \boldsymbol{E}_N) d\Omega$ is formally related to the quantum correlated projection $(\langle \boldsymbol{e}_1|\langle \boldsymbol{e}_2| \ldots \langle \boldsymbol{e}_N|)|\psi_N\rangle$. The unit vector $\boldsymbol{e}_i$ ($i = 1, \ldots, N$) denotes the direction of

the polarization projection of the beam $E_i$, and $|e_i\rangle$ is the corresponding projection state of the $i$th qubit[40]. A detailed explanation of the correspondence between beam states and qubit states is given in the first section of the Materials and methods. Based on Eq. (1), a data sample expressed by a $2^N$-dimensional complex vector can be encoded by the correlations of above $N$ beams. In what follows, we show that encoding the data by the correlations enables a better way to process it.

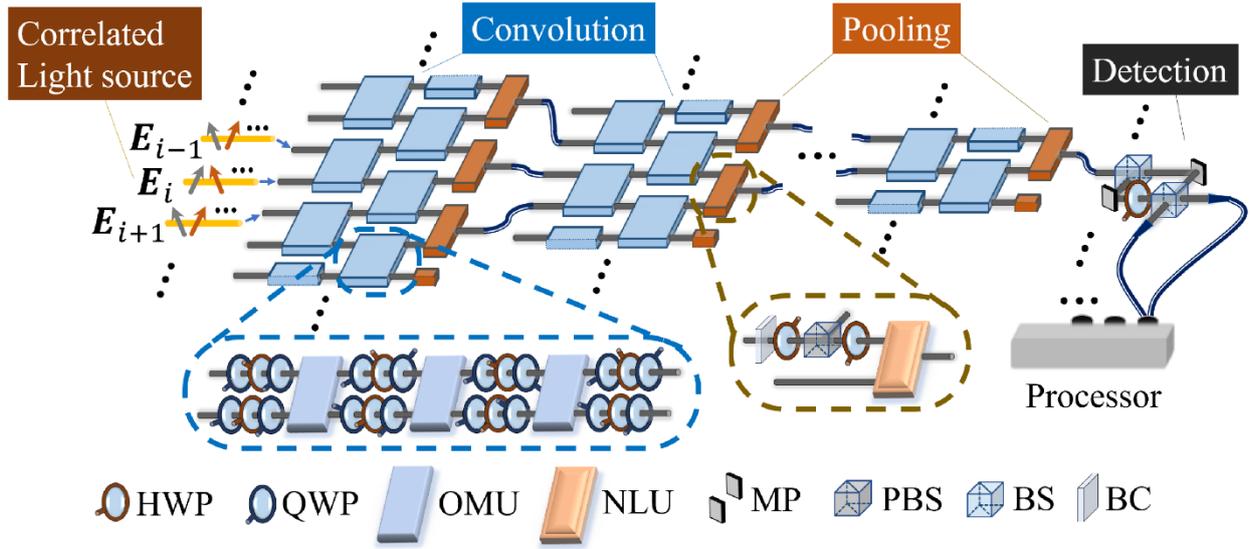

**Fig. 1 |** The general setup of a COCNN. The correlated light source, whose polarized modes are marked by arrows with different colors, is manipulated by the parts of the convolution (blue) and the pooling (brown). The part of the convolution is composed of the 2-beam operations marked by blue, and the details of a 2-beam operation is given in the blue dashed box. As shown inside the box, a 2-beam operation is implemented by eight Q-H-Qs and three NLUs. The part of the pooling is composed of the combiners marked by brown, and the details of a combiner is given in the brown dashed box. As shown inside the box, a combiner is implemented by a BC, an HWP-PBS-HWP, and an NLU. Finally, the detection of the output is performed by a homodyne interferometer and a processor. The interferometer is mainly built by a pair of mirrors (MP) and two beam splitters (BSs).

Secondly, we introduce the part of the convolution, as shown by the layers of the connected blue squares in Fig. 1. This is an essential part for processing the encoded data in our proposal. Traditionally, the convolution of the data is mathematically expressed by the linear transformation of a vector, which is the main function of the convolution blocks in the previous ONNs[5,9,18]. Here, the part of the convolution is also the linear transformation, but applied to the correlation rather than the amplitude of the beams of light. This leads to a completely different way of design for implementing the transformation. Following Eq. (1), the convolution can be described by a unitary operation $U_c$ on $|NE)$. Here, we restrict that $U_c$ is composed of general 2-beam operations $U_{2E}$, as shown by the blue squares in Fig. 1. Given a state of the two correlated beams $|2E) = c_{h_1 h_2} |h_1)|h_2) + c_{h_1 v_2} |h_1)|v_2) + c_{v_1 h_2} |v_1)|h_2) + c_{v_1 v_2} |v_1)|v_2)$, $U_{2E}$ represents the operations that can rotate $|2E)$ to an arbitrary state in the space spanned by $\{|h_1)|h_2), |h_1)|v_2), |v_1)|h_2), |v_1)|v_2)\}$. It is the analogy of the universal 2-qubit gate in the quantum computing theory. The recipe for implementing $U_{2E}$ is shown in the blue dashed box in Fig. 1[40], including two basic arrangements of the optical elements. The first is the Q-H-Q, composed of two quarter-wave plates (QWP) and one half-wave plate (HWP). A Q-H-Q can arbitrarily rotate the polarization state of a single beam, which is equivalent to the rotation of a single qubit. The second is an optical modulation unit (OMU), shifting the phase of the state component $|v_1)|v_2)$ by a factor of $\pi$. To realize such an element is actually tricky, and we provide one kind of design in S1 of the Supplementary Information. The function of an OMU is equivalent to the quantum CZ gate. The reason why the setup of the Q-H-Qs and the OMUs in Fig. 1 implements an arbitrary rotation of $|2E)$ can be explained in refereeing to the quantum computing theory. In the theory of quantum computing, a universal 2-qubit gate can be decomposed into three CNOT gates and eight single qubit rotations[40], and a CNOT gate can be synthesized by a CZ gate and two Hadamard gates[41]. Therefore, the decomposition of $U_{2E}$ into Q-H-Qs and OMUs can be correspondingly given. According to the definition of a $U_{2E}$, it can be parameterized by 15 real numbers, which is the same with the parameter number of a 2-qubit gate. So, as shown by Fig. 1, one layer of $U_{2E}$ acting on the adjacent beams of an $N$-beam array may have $15 \times (N-1)$ trainable parameters in total. This is larger than the convolution kernel applied in the standard convolutional neural networks (CNNs) framework[42], or other equivalent ONN schemes[5,9,18], which has at most $N$ parameters for an $N$-dimensional input.

Thirdly, we introduce the part of the pooling, as shown by the layers of the brown squares in Fig. 1 with two inputs and one output. This is the key step for reducing the size of the data. Particularly, each brown square is called a combiner whose basic function is to generate simpler correlated states by decreasing the beam number. The function of the combiner is strictly given by

$$|2E)(2E| \rightarrow [(h_1| \otimes I]|2E)(2E|[|h_1) \otimes I] + [(v_1| \otimes I_2]|2E)(2E|[|v_1) \otimes I] \quad (2)$$

where $(2E|$ $((h_1|, (v_1|)$ is the Hermitian conjugate of $|2E)$ $((h_1|, (v_1|)$. The form $|2E)(2E|$ is analog to the density matrix of a quantum state, and we also call it the density matrix of the beam state $|2E)$. $I$ is the identity operation on a single beam state. Notice that Eq. (2) is analog to the partial trace in quantum 2-partite system, which physically means looking into the single particle subspace of the system. Therefore, via the combiner defined by Eq. (2), the correlated state of the two beams is embedded into the state of one beam, losing part of the original information. Hence, if the combiner is applied for $k$ times, an array of $N$ beams is reduced to an array of $N-k$ beams. The recipe of implementing Eq. (2) is shown in the brown dashed box in Fig. 1. At first, the upper beam is fed into a birefringence crystal (BC). The thickness of the BC is demanded to break the coherence of the horizontal and vertical polarization components. Then, the beam passes through an element set of HWP-PBS-HWP. Such an element set is equivalent to a polarizer, which can change an arbitrary state $\alpha \boldsymbol{h} + \beta \boldsymbol{v}$ to $(\alpha + \beta)(\boldsymbol{h} + \boldsymbol{v})/2$. Then, the beam is modulated onto the lower beam through a nonlinear crystal unit (NLU). The basic function of an NLU is to generate $[\epsilon(\alpha + \beta)/2]\boldsymbol{h} + [\gamma(\alpha + \beta)/2]\boldsymbol{v}$ with the input $(\alpha + \beta)(\boldsymbol{h} + \boldsymbol{v})/2$ and another beam $\epsilon \boldsymbol{h} + \gamma \boldsymbol{v}$, where $\epsilon$ and $\gamma$ are also complex coefficients of the polarization basis. The function of an NLU means that the $\boldsymbol{h}$ (or $\boldsymbol{v}$) component of one mode in the output beam is the product of the $\boldsymbol{h}$ (or $\boldsymbol{v}$) components of the corresponding modes of the input beams. This is generally the second-order nonlinear crystal in nature. A detailed analysis is given in the S2 of the Supplementary Information. Notice that by using one layer of the combiner, the information of the data encoded in the $N$-beam correlation is suppressed into those of $N-k$ beams. It indicates that the dimension of the state space shrinks from $2^N$ to $2^{N-k}$, which is more efficient than the layer for average pooling or max pooling in the traditional CNNs, or other ONN models involving equivalent structures[9].

Fourthly, we introduce the part the of the detection. The basic aim of the detection is to measure the correlation defined by $\int (\boldsymbol{e}_1 \cdot \boldsymbol{E}_1)(\boldsymbol{e}_2 \cdot \boldsymbol{E}_2) \dots (\boldsymbol{e}_N \cdot \boldsymbol{E}_N) d\Omega$. In our proposal, we consider a two-step procedure to achieve so. In the first step, each beam interferes with a local oscillator (LO) signal by using a homodyne interferometer, so that the projection of the $i$th beam $\boldsymbol{e}_i \cdot \boldsymbol{E}_i$ can be obtained by the intensity difference of the outputs of the interferometer. The polarization of LO signal for measuring the $i$th beam $\boldsymbol{E}_i$ is set to be at the projection direction $\boldsymbol{e}_i$. The LO signal is coherent with all modes of $\boldsymbol{E}_i$, and can be generated by splitting the beam of $\boldsymbol{E}_i$, as shown by the rightmost part of Fig. 1. In the second step, all the projections $\boldsymbol{e}_i \cdot \boldsymbol{E}_i$ $(i = 1, \dots, N)$ measured by the homodyne detections are multiplied together by a processor. Then, the obtained result is proportional to the correlation defined by the above integral. In fact, the detection setup here is identical to the one employed in Ref. (40). The underlying background of the setup is also the correspondence between the beam states defined by Eq. (1) and the qubit states, which is given in the first section of the Materials and methods as mentioned above.

The training of a COCNN is similar to other machine learning algorithms. By properly setting the loss function, one can assess the performance of the COCNN on a given dataset and update the parameters according to it. Here, we also employ the mean square error (MSE) as the loss function. If the target output of the $i$th data sample is denoted by $y_i$, and the corresponding output of the COCNN is denoted by $y'_i$, the MSE can be given by

$$MSE = \frac{1}{2D}\sum_{i=1}^{D}(y_i - y_i')^2 \tag{3}$$

where $D$ is the total number of the samples.

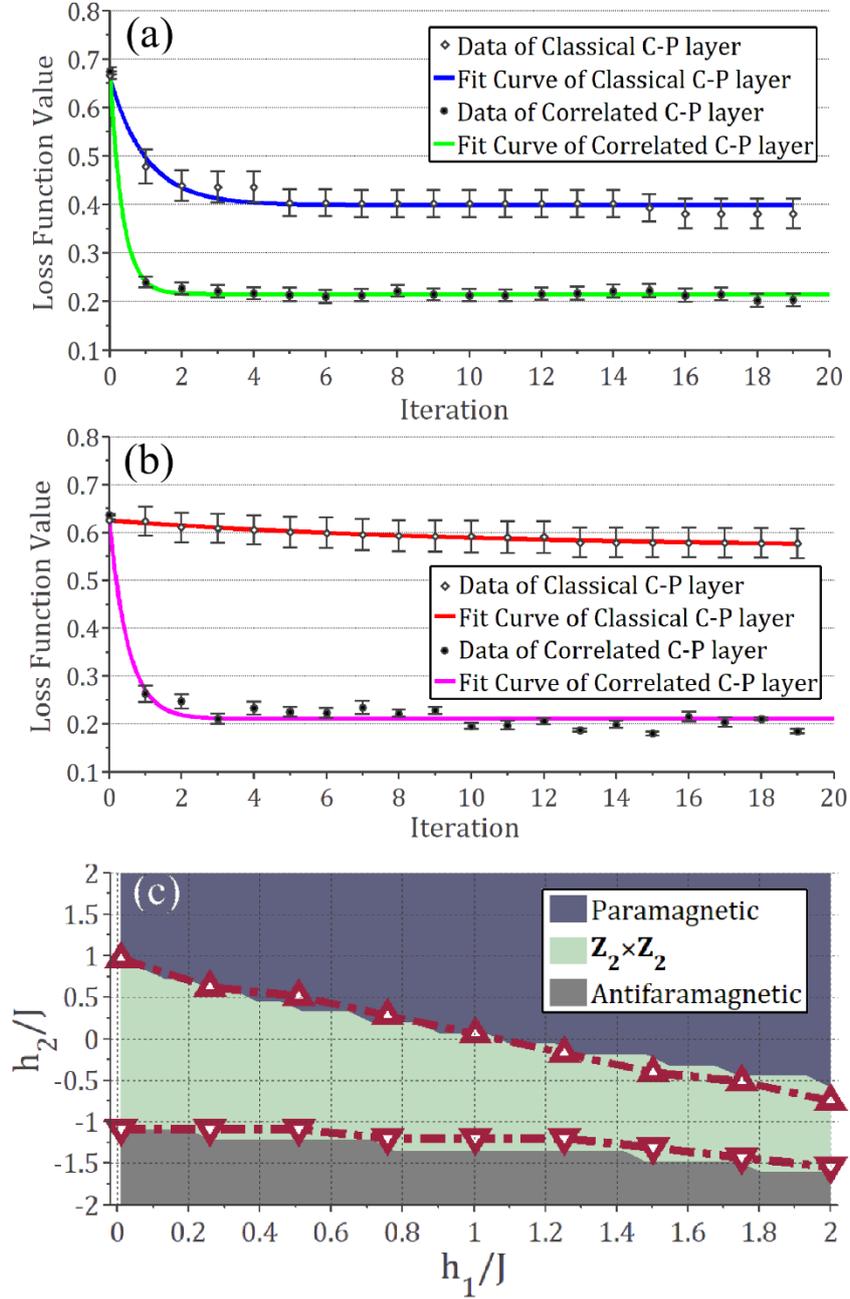

**Fig. 2** | (a) The comparison between the training of a COCNN model and that of a traditional CNN model on a binary classification task, and (b) that on a multiclass classification task. In (a) and (b), the hollow squares and solid dots are the numerical results. The blue curve in (a) and the red curve in (b) are fitted by the squares obtained by the traditional CNN model, which is composed of a traditional convolutional layer and a traditional pooling layer (the classical C-P layer model). The green curve in (a)

and the magenta curve in (b) are fitted by the dots obtained by the COCNN model, which is composed of a convolutional layer and a pooling layer in our scheme (the correlated C-P layer model). The error bar is marked on the corresponding dots. (c) The simulation of the 12-qubit QCNN ansatz for recognizing the Haldane phase by our COCNN. The background of the plot is colored according to the derivative of the ground state energy density of the Haldane Hamiltonian. The upper zone represents the paramagnetic phase. The middle zone represents the $Z_2 \times Z_2$ phase, or the Haldane phase. The lower zone represents the anti-paramagnetic phase. The red triangles are obtained by the output of the COCNN analog to the QCNN.

According to the above description, it can be noticed that the COCNN has a good correspondence with the QCNN proposed by Ref. (37). This can be seen by the similarity between the correlation $\int (\boldsymbol{e}_1 \cdot \boldsymbol{E}_1)(\boldsymbol{e}_2 \cdot \boldsymbol{E}_2) \ldots (\boldsymbol{e}_N \cdot \boldsymbol{E}_N) d\Omega$ and the quantum measurement $(\langle \boldsymbol{e}_1 | \langle \boldsymbol{e}_2 | \ldots \langle \boldsymbol{e}_N |) | \psi_N \rangle$. Such a relation indicates that the COCNN can exhibit the properties of the QCNN accordingly. We show the relevant evidence from the following two aspects. The first one is about the speed-up in training. We numerically explore the training process of a COCNN model on two datasets, and take the performance of the CNN model on the same tasks as reference. In general, the considered tasks are to give the correct label of the input data sample by training the model parameters. The results are shown in Fig. 2(a) and 2(b). The results in Fig. 2(a) are the convergence curves of the loss function for a binary classification task. The dataset we consider is composed of 0-1 vectors whose sizes are $8 \times 1$. There are 8 types of vectors in the set, each of which is labeled by 0 or 1. The distribution of the vectors is uniform, and each type of the vectors is randomly labeled. In the COCNN scheme, it only requires the three-beam states, $|3E\rangle = c_{h_1 h_2 h_3} |\boldsymbol{h}_1\rangle |\boldsymbol{h}_2\rangle |\boldsymbol{h}_3\rangle + c_{h_1 h_2 v_3} |\boldsymbol{h}_1\rangle |\boldsymbol{h}_2\rangle |\boldsymbol{v}_3\rangle + \cdots + c_{v_1 v_2 v_3} |\boldsymbol{v}_1\rangle |\boldsymbol{v}_2\rangle |\boldsymbol{v}_3\rangle)$, to encode this kind of data samples according to our setup. The COCNN model for the task is composed of one convolutional layer (C-layer) and one pooling layer (P-layer). The C-layer contains two 2-beam operation $U_{2E}$s, each of which is restricted to has 6 real trainable parameters. The P-layer contains two combiners, leaving only one beam as the output. The predicted label is given by the projection of the output beam, and the projective measurement is parameterized by 3 real numbers (two for the direction and one for the phase). By sequentially updating the 15 parameters based on the derivative of the loss function, the model finally converges. To verify the robustness of the results, we randomly generate the training dataset for 12 times and take the averages as the final results, as shown by the solid dots in the lower place of Fig. 2(a). The green curve is fitted by this part of the results. As the reference, we establish a 15-parameter CNN to learn the same series of datasets, and adopt the MSE as the loss function as well. We also utilize the averages of the 12 trails as the results, as shown by the hollow squares in the upper place of Fig. 2(a). The blue curve is fitted by this part of the results. Apart from the case of Fig. 2(a), the COCNN can also be applied to complicated multiclass classification tasks. In order to show the point, we also consider the task on classifying a dataset containing four classes, shown in Fig. 2(b). The dataset we consider is composed of 0-1 vectors whose sizes are $16 \times 1$. There are one hundred types vectors in the set, each of which is randomly labeled by 0, 1, 2, or 3. The distribution is approximately uniform. Using the COCNN scheme, the four-beam states, $|4E\rangle = c_{h_1 h_2 h_3 h_4}|h_1\rangle|h_2\rangle|h_3\rangle|h_4\rangle + \cdots + c_{v_1 v_2 v_3 v_4}|v_1\rangle|v_2\rangle|v_3\rangle|v_4\rangle$, are employed to encode the data samples. The COCNN model for the task is also composed of one C-layer and one P-layer. The C-layer contains three 2-beam operation $U_{2E}$s, each of which is also restricted to has 6 real trainable parameters as well. The P-layer contains two combiners, leaving two beams as the output. The predicted label is given by the correlated projection of the two output beams, each local measurement of which is parameterized by 3 real numbers as the above case. The total parameter number is 26. We also generate the training data set for 12 times, and take the averages to obtain the solid dots in the lower place of Fig. 2(b). The magenta curve is fitted by this part of the results. The reference results shown by the red curve are fitted by the loss function value (marked by hollow squares in Fig. 2(b)) of a 26-parameter CNN, and also averaged over 12 trails. The loss function is the same with the above. The details are presented in the second section of the Materials and methods. It can be seen clearly from Fig. 2(a) and 2(b) that the COCNN model converges faster than the CNN model. Also, the loss function of the COCNN model eventually converges to a smaller value than that of the CNN model, indicating a better learning accuracy. It worth noticing that the performance revealed by Fig. 2(a) and 2(b) is comparable with the numerical results in Ref. (39). Therefore, the speed-up we find here is as effective as that of a QCNN.

The second one is about performing the specific function of a QCNN. As pointed out by Ref. (37), one function of a QCNN is to identify the phase of a many-body quantum system. We consider the QCNN circuit for recognizing the Haldane phase. As mentioned above, a correlated projection of a quantum state can be mapped to the correlation of the beams. So, the ground state of the Haldane Hamiltonian,

$$H = -J \sum_{i=1}^{N-2} Z_i X_{i+1} Z_{i+2} - h_1 \sum_{i=1}^{N} X_i - h_2 \sum_{i=1}^{N} X_i X_{i+1} \tag{4}$$

can also be encoded by the correlated beams we consider. $X_i$ and $Z_i$ in Eq. (4) are the Pauli-$X$ and Pauli-$Z$ operator on the $i$th particle. Integer $N$ is the total number of the particles, and $J$, $h_1$ and $h_2$ are the coefficients of the Hamiltonian[43]. In Ref. (37), a strict QCNN is presented to justify the phase of the ground states under different $J$, $h_1$ and $h_2$. Because the convolution and pooling of our scheme have a good correspondence with the quantum counterparts in QCNN, our COCNN can also implement the phase recognition algorithm based on the QCNN. An instruction about the detailed setup is given in the second section of the Materials and methods. We numerically simulate the situation when $N = 12$ and the results are shown in Fig. 2(c). The phases identified with the ground state energy density is marked by the color of the background as the reference. The red triangles in Fig. 2(c) denotes the phase boundary information obtained by the second order derivative of the simulated results of the COCNN for phase recognition. Corresponding to the QCNN scheme, the results of the COCNN are the projections of the output beam states on the Pauli-$X$ basis of the polarization. Notably, due to the layers of pooling, the correlation of the output is involved by fewer beams than the input ones encoding the ground state. Hence, such an output contains the information for identifying the phase and is much easier to measure, just like the characters shown by the QCNN. In the particular case here, the phase of the state encoded by the correlation of the 12 beams is eventually recognized by measuring the correlation of the 3 output beams. It also worth noticing that the results in Fig. 2(c) is comparable with the numerical illustration in Ref. (37), validating the correspondence between the COCNN and the QCNN. Therefore, our scheme is capable of carrying out data processing similar to that of a quantum network. We also provide a detailed analysis of the connections between COCNNs and QCNNs in S3 of the Supplementary Information.

## The experimental realization of a COCNN

In this part, we explore the experimental demonstration of the above scheme. Taking the COCNN analog to the 3-qubit phase recognition QCNN as an example, we show our experimental setup in Fig.3. The setup also performs the functions explained in Fig. 1, including the part of the correlated light source, convolution, pooling, and measurements. The part of correlated light source in the experimental setup is implemented by adopting the spatial modes as the $f_k$ of the beam state. The spatial modes are generated by one 632.8nm He-Ne laser (ThrolabsHNL210LB). The light output by the laser is polarized by a beam displacer (BD) and then split into four spatial modes by three beam splitters (BS, ThorlabsBS016). For each mode, the polarization state is adjusted by a Q-H-Q. The filters thereafter are used for balancing the intensity of different modes. This is a simplified version of the strategy for generating the correlated light source in the above discussions. The simplification is based on the intrinsic relation between the beam state defined by Eq. (1) and the corresponding qubit states. As mentioned above, the intrinsic relation is strictly characterized by the Eqs. (S6) in S1 of the Supplementary Information, indicating that there could be more than one setup for the beams in order to mimic a given quantum state. Therefore, it can be explored to identify the most experiment-friendly beam setup among the alternatives. In our case, we choose to settle down the polarization components of the two beams of the set, and implement the required operation effectively by manipulating the states of the rest beam. Hence, the analogy of the quantum circuit can be realized by only one beam involving four spatial modes, as shown by the left of Fig. 3. This largely reduces the requirements on the experiments. A strict explanation of the simplification is presented in the third section of the Materials and methods.

Next, the modes are fed into the part of convolution, corresponding to the blue layers in Fig. 1. The first operation in this part is an operation analog to the two CZ gates on adjacent qubits. After simplification, the operation is implemented by the pair of HWPs in the left yellow region. The one on the second mode shifts the phase of the vertical component by a factor of $\pi$, and the one on the third mode shifts the phase of the same component by the same magnitude. The second operation is a combination of the two single-beam rotations, which is analog to the two quantum Hadamard gates on the first and the third qubit respectively. In the general proposal, it can be realized by two Q-H-Qs. Here, it is implemented by the four BSs in the gray region. The upper left BS mix the first and the third modes, and the lower left BS mix the second and the fourth modes. The right two BSs mix the outputs of the left ones. The third operation is a three-beam operation, which is the analogy of quantum Toffoli gate with the second qubit serving as the target qubit. According to our general proposal, it can be realized by using a series of 2-beam operations. Through the simplification strategy here, it is implemented by the HWP in the orange region. This HWP exchange the horizontal and vertical component of the fourth mode. The fourth operation is the same with the first operation. Therefore, it is also implemented by a pair of HWPs whose area is marked by yellow. The fifth operation is a single beam rotation, analog to the single Hadamard gate on the second qubit. This operation is applied for changing the basis of the projection in the next part. Here, it is implemented by four HWPs in the red region. Each HWP in the region changes the $\boldsymbol{h}$ polarization component to $(\boldsymbol{h} + \boldsymbol{v})/\sqrt{2}$ and changes the $\boldsymbol{v}$ polarization component to $(\boldsymbol{h} - \boldsymbol{v})/\sqrt{2}$. Then, the following projection of the beam is going to be casted in the Pauli-$X$ basis.

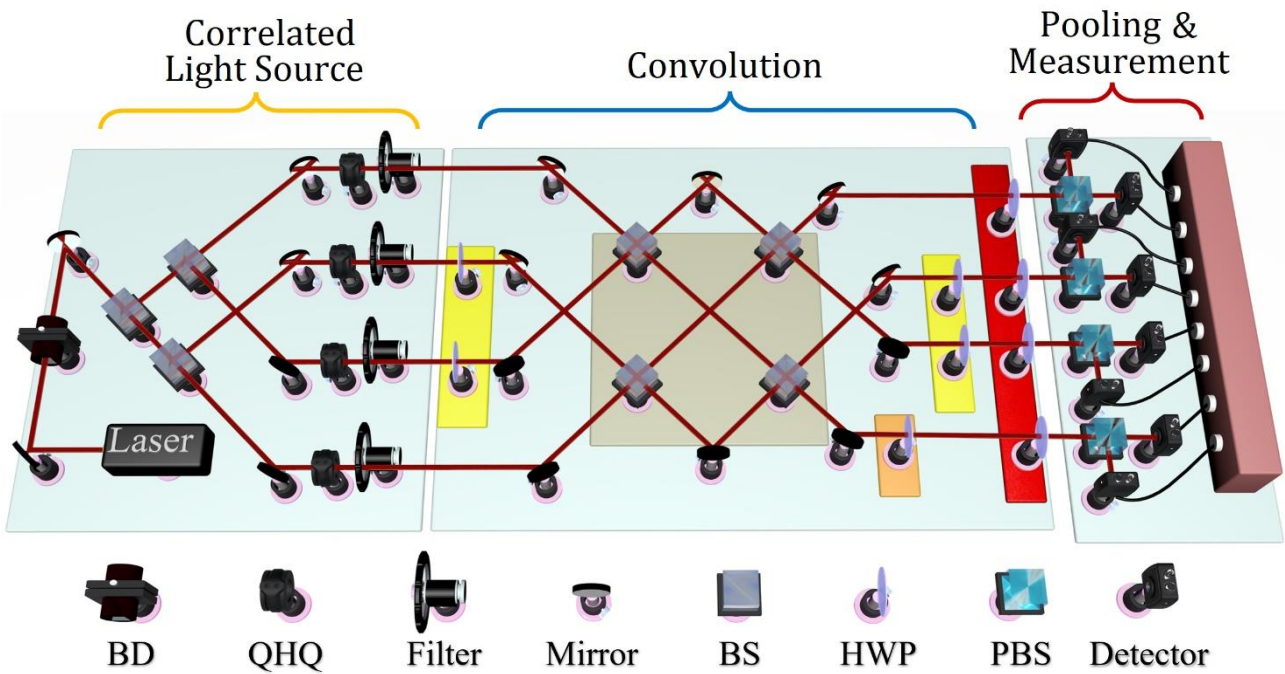

**Fig. 3** | The experimental realization of a simplified COCNN analog to the 3-qubit QCNN circuit for the

phase recognition. The pooling and the detection part are combined into one due to the simplification. The optical elements used in the experiment is listed below. The related elements for performing an operation are marked by the area of a certain color.

After the part of convolution, the spatial modes go into the region for pooling and measurements, corresponding to the brown layers and the detections in Fig. 1. Because of the simplification we consider, the pooling operation in the experiment is realized by measuring the sum of the intensity difference of all the modes (see S4 of the Supplementary Information for details). Hence, it is merged into the measurements as shown in Fig. 3. In the region, the polarization of each mode is divided by a polarized beam splitter (PBS), and finally recorded by the CCDs (Thorlabs BC106N-VIS/M). By using such a detection setup, one can obtain the intensity sum of the vertical or horizontal components of the output. Further, the results corresponding to the quantum measurements can be acquired. Suppose the output state of our optical setup is denoted by $|E_{out}\rangle$. Adding up the intensities of all the components gives the result $\text{Tr}\{|E_{out}\rangle\langle E_{out}|\}$, and subtracting the vertical component sum from the horizontal component sum gives the result $\text{Tr}\{Z|E_{out}\rangle\langle E_{out}|\}$, where $\text{Tr}\{\ \}$ means taking the trace. If one arranges an HWP or a QWP oriented at 22.5° before each PBS in the last region, the results $\text{Tr}\{X|E_{out}\rangle\langle E_{out}|\}$ or $\text{Tr}\{Y|E_{out}\rangle\langle E_{out}|\}$ can also be obtained by the subtraction, where $X$, $Y$, and $Z$ are the Pauli operators. These results correspond to the quantum measurements $\text{Tr}\{\rho_{out}\}$, $\text{Tr}\{Z\rho_{out}\}$, $\text{Tr}\{X\rho_{out}\}$ and $\text{Tr}\{Y\rho_{out}\}$ respectively, where $\rho_{out}$ is the output state of a quantum circuit. Similar to the strategy for estimating a single qubit state, the estimation of the single beam state $|E_{out}\rangle\langle E_{out}|$ can be given by $I \cdot \text{Tr}\{|E_{out}\rangle\langle E_{out}|\} + X \cdot \text{Tr}\{X|E_{out}\rangle\langle E_{out}|\} + Y \cdot \text{Tr}\{Y|E_{out}\rangle\langle E_{out}|\} + Z \cdot \text{Tr}\{Z|E_{out}\rangle\langle E_{out}|\}$. Notice that function of our experimental setup also has a one-to-one correspondence with the 3-qubit phase recognition QCNN[37]. The 3-qubit circuit is shown by Fig. S3 in S4 of the Supplementary Information, composed of four CZ gates, three Hadamard gates and one Toffoli gate. The final output of the circuit is given by the measurements on a single qubit state. A detailed instruction of the whole theoretical background is provided in S4 of the Supplementary Information.

To benchmark the performance, we firstly check the output of the above optical setup when the inputs are the analogies of several special quantum states. Particularly, we consider the ten states of the beams, including $|h_1\rangle|h_2\rangle|h_3\rangle$, $|v_1\rangle|h_2\rangle|h_3\rangle$, $|v_1\rangle|v_2\rangle|h_3\rangle$, $|v_1\rangle|v_2\rangle|v_3\rangle$, $|p_1\rangle|m_2\rangle|p_3\rangle$, $|m_1\rangle|p_2\rangle|m_3\rangle$, $|l_1\rangle|r_2\rangle|l_3\rangle$, $|r_1\rangle|l_2\rangle|r_3\rangle$, $[|h_1\rangle|h_2\rangle|h_3\rangle + |v_1\rangle|v_2\rangle|v_3\rangle]/\sqrt{2}$, and $[|h_1\rangle|h_2\rangle|v_3\rangle + |h_1\rangle|v_2\rangle|h_3\rangle + |v_1\rangle|h_2\rangle|h_3\rangle]/\sqrt{3}$. $|p\rangle$ and $|m\rangle$ denote $[|h\rangle + |v\rangle]/\sqrt{2}$ and $[|h\rangle - |v\rangle]/\sqrt{2}$ respectively. $|r\rangle$ and $|l\rangle$ denote $[|h\rangle + i|v\rangle]/\sqrt{2}$ and $[|h\rangle - i|v\rangle]/\sqrt{2}$ respectively. The corresponding ten quantum states are $|000\rangle$, $|100\rangle$, $|110\rangle$, $|111\rangle$, $|+-+\rangle$, $|-+-\rangle$, $|LRL\rangle$, $|RLR\rangle$, $|GHZ\rangle = (|000\rangle + |111\rangle)/\sqrt{2}$ and $|W\rangle = (|001\rangle + |010\rangle + |100\rangle)/\sqrt{3}$, where $|\pm\rangle$ denotes $(|0\rangle \pm |1\rangle)/\sqrt{2}$, and $|R\rangle$ ($|L\rangle$) denotes $(|0\rangle + i|1\rangle)/\sqrt{2}$ ($(|0\rangle - i|1\rangle)/\sqrt{2}$). The theoretical and experimental results of $|E_{out}\rangle\langle E_{out}|$ under different inputs are shown from Fig. 4(a) to 4(j), in the order of the above ten states. The fidelities of the results are 0.997, 0.9998, 0.9995, 0.9996, 0.9985, 0.9840, 0.9986, 0.9989, 0.9853, and 0.9988, respectively. The theoretical calculation of the experimental output for the ten input states is given in S5 of the Supplementary Information.

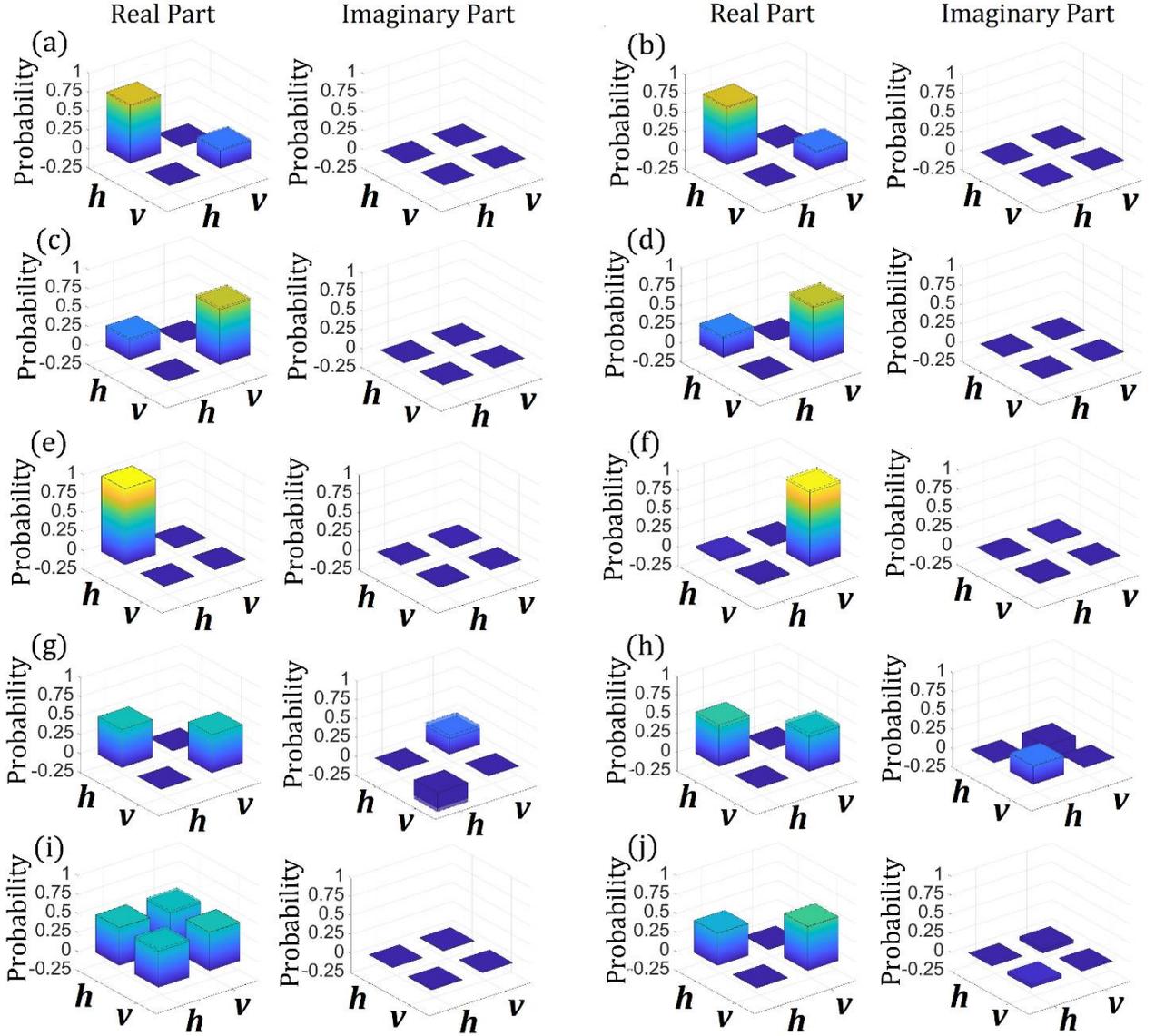

**Fig. 4** | The density matrices of the COCNN outputs. The results are obtained by measuring the quantities $\mathrm{Tr}\{|E_{out}\rangle\langle E_{out}|\}$, $\mathrm{Tr}\{X|E_{out}\rangle\langle E_{out}|\}$, $\mathrm{Tr}\{Y|E_{out}\rangle\langle E_{out}|\}$, and $\mathrm{Tr}\{Z|E_{out}\rangle\langle E_{out}|\}$. The input beam states of (a)-(j) are analog to the qubit states $|000\rangle$, $|100\rangle$, $|110\rangle$, $|111\rangle$, $|+-+\rangle$, $|-+-\rangle$, $|LRL\rangle$, $|RLR\rangle$, $|GHZ\rangle$, and $|W\rangle$ respectively. The heights of the colored inertia are the experimental data, and the heights of the black frame of the cuboids are the theoretical data. The theoretical expression of the results can be found in S5 of the Supplementary Information.

Next, we present the phase recognition results. The input states here are set to be the analogies of the ground states of the 3-site Haldane Hamiltonian. The basic strategy to encode the ground states into the beams is given in S6 of the Supplementary Information. The ground states are calculated by the diagonalization. Here, $h_1/J$ is set to be 0.4, 0.8, 1.2 and 1.6, and $h_2/J$ is taken from -2.0 to 2.0 with an interval of 0.25. According to the numerical simulation of the COCNN scheme, we measure $\text{Tr}\{Z|E_{out})(E_{out}|\}$ in this case, which is equivalent to the measurement of the QCNN phase recognition circuit due to the basis transformation enabled by the four HWPs in the red region. The results are marked by the red dots in the left panel of Fig. 5, and the data for obtaining the results is provided in S6 of the Supplementary Materials.

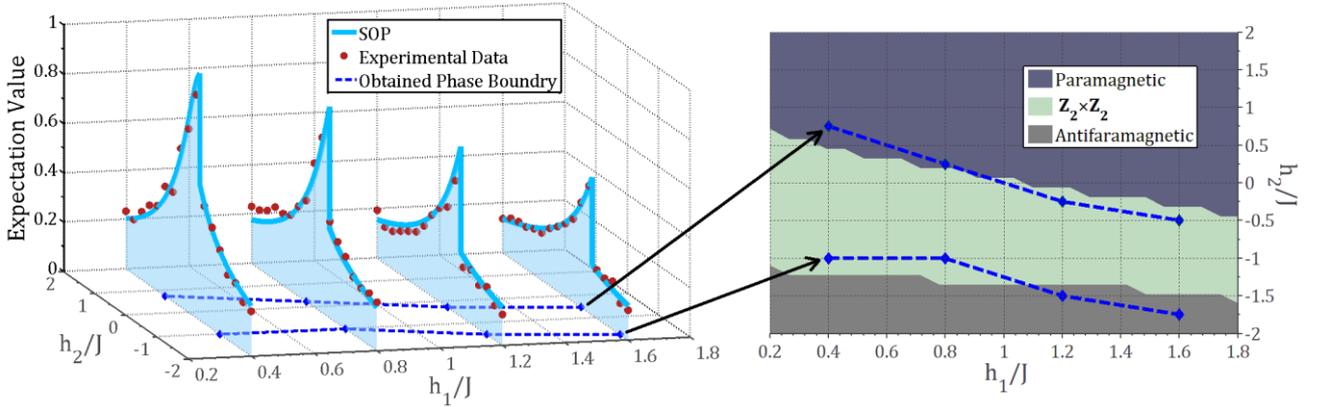

**Fig. 5** | The results when the input states encode the Haldane ground states. The red dots in the left panel are the experimental data. The *X*- and *Y*-axis of the coordinates represent the ratios $h_1/J$ and $h_2/J$. The *Z*-axis represent the expectation $\text{Tr}\{Z|E_{out})(E_{out}|\}$ in our basis, equivalent to the measurements in the QCNN phase recognition circuit. The light blue curves are obtained by the SOP. The dark blue curves are the phase boundaries obtained by the second order derivative of the experimental data. The right panel display the comparison of the phase boundaries obtained by the experimental data and those shown in Fig. 2(c).

For comparison, we plot the standard phase recognition results obtained by string-order parameters (SOP)[37], shown by the light blue curves in left panel of Fig. 5. In the 3-site case, the SOP is given by $\langle ZXZ \rangle$. From the left panel, we can see that the experimental data fits well with the SOP results, validating the effectiveness of the setup. Besides, the phase boundaries can also be obtained by the second order derivative of the experimental data, shown by the dark blue curves in Fig. 5. Compared with the numerical simulation in Fig. 2(c), it can be found that the dark blue curves match the boundaries obtained by the numerical results. A direct illustration is shown by the right panel of Fig. 5. It worth noticing that the original SOP is a quantity that requires to measure the three-particle correlations. Using the COCNN here, it is effectively reduced to measuring the character of a single particle. This phenomenon in the COCNN experiment reveals the benefit of applying the QCNN for recognizing Haldane phase proposed by Ref. (37).

Additionally, it can be found that the 3-qubit phase recognition QCNN functions as the fundamental building block of the $N$-qubit phase recognition QCNN due to the repeated structure of the network (as shown in Fig. S5 of the Supplementary Information). This implies that the experimental setup of the COCNN, which performs the same operation as the 3-qubit phase recognition QCNN, can also be viewed as the fundamental building block of the phase recognition COCNN that utilizes $N$-beam states. Moreover, for implementing other complicated tasks, the parameters of our general COCNN explained in the second section will be trained on the specific datasets of the tasks. The setups for those tasks can also be given by using the similar arrangements as in Fig. 3, according to our general proposal and the first section of the Materials and methods. Meanwhile, the simplification strategy applied here can also be extended to those tasks involving $N$ beams. Thus, in principle, the $N$-beam COCNNs for the task or others are also implementable.

## Discussion

In summary, we have proposed to introduce the correlation of the light fields for establishing a new ONN framework, which is called as the COCNN. Different from the previous ONN, which only adopts the superposition property of light, the COCNN can exhibit similar characters of the quantum neural networks. We numerically show that for the classification tasks we consider, the loss function of a COCNN converges faster than that of a CNN. Moreover, we have also shown the COCNN can be applied to implement the function of the QCNNs, such as the one for the recognition of the Haldane phase. Considering the fact that the COCNN we propose has a one-to-one correspondence with the quantum circuit, the speed-up here could be as effective as the speed-up of a QCNN. Taking the COCNN analog to the 3-qubit phase recognition QCNN as an example, we have explored the experimental demonstration of the COCNN. The function of the setup has been firstly checked by setting the input to be the analogy of ten quantum states, and secondly set to perform the phase recognition for the ground states of the 3-site Haldane Hamiltonian. All the experimental results are in good agreement with the theoretical results of the QCNN, indicating that the function of the QCNN can be realized by using our COCNN scheme in principle.

As mentioned above, the main character of the COCNN strategy is the modulation of the correlated beam states. It is the major cause of the acceleration in the training process. On the basis of the modulation, a convolutional operation can be given, enabling a quite effective capture of the data feature. More importantly, the pooling operation in this manner can reduce the size of the processed data faster than the traditional pooling of the CNN. The two operations of the COCNN are equivalent to those applied in the QCNN. Hence, we think that such a scheme potentially advances the boundaries of optical acceleration. Meanwhile, the results also indicate that the

COCNN allows for the realization of the properties of quantum neural network in a more affordable way. Despite the potential advantages of quantum neural networks, implementing them practically requires deep quantum circuits with many multi-qubit gates and complicated measurements. This necessitates significant resources to stabilize the circuits and correct errors, which is technically challenging due to the unavoidable environmental disturbances. A potentially better alternative to implementing advantageous algorithms suggested by quantum computing theory is to find a system described by the same math as quantum theory and interrupted less by the environment. The proposed COCNN serves as an example of such a system, as evidenced by the ease of element arrangements and low requirements on the circumstances in our experiments. In all, given the exponential growth of data and the scarcity of resources for high-quality computation, the COCNN we propose presents a cost-effective and high-performance solution that could have widespread applications in various data science research fields.

## Materials and methods

### The correspondence between beam states and qubit states

The array of the beams we consider has been introduced in Ref. (40). Here we briefly review the basic setup. Consider the 2-beam state $|2E\rangle = c_{h_1 h_2} |h_1\rangle|h_2\rangle + c_{h_1 v_2} |h_1\rangle|v_2\rangle + c_{v_1 h_2} |v_1\rangle|h_2\rangle + c_{v_1 v_2} |v_1\rangle|v_2\rangle$ as an example. The state can be given by two beams

$$\boldsymbol{E}_1 = \sum_{k=1}^{M} f_k \boldsymbol{p}_{1,k}, \qquad \boldsymbol{E}_2 = \sum_{k=1}^{M} f_k \boldsymbol{p}_{2,k} \tag{5}$$

As we mentioned in the main text, the LO beam for measuring $\boldsymbol{E}_1$ and $\boldsymbol{E}_2$ are expressed by $\boldsymbol{E}_1^{\text{LO}} = F\boldsymbol{e}_1$ and $\boldsymbol{E}_2^{\text{LO}} = F\boldsymbol{e}_2$. $F$ represents the mode coherent with all $f_k$s, such that $F \cdot f_k \propto f_k$. Using the homodyne detection, one obtains the real part of the projection of $\boldsymbol{E}_1$

$$\text{Re}\{D_1\} = |\boldsymbol{E}_1 + \boldsymbol{E}_1^{\text{LO}}|^2 - |\boldsymbol{E}_1 - \boldsymbol{E}_1^{\text{LO}}|^2 = 2\ \text{Re}\{\boldsymbol{E}_1^{\text{LO}*} \cdot \boldsymbol{E}_1\} \tag{6}$$

as well as that of $\boldsymbol{E}_2$ denoted by $\text{Re}\{D_2\}$. $\text{Re}\{\ \}$ means taking the real part. The imaginary part can be obtained by shifting the phase of $\boldsymbol{E}_1^{\text{LO}}$ and $\boldsymbol{E}_2^{\text{LO}}$ by $\pi/4$. Then, one can obtain the correlation by multiplying $D_i$ and $D_{i+1}$ and integrating the product in the domain where the orthonormal condition of $f_k$ holds. This is given by $\int D_1 D_2 \, d\Omega \propto \int (\boldsymbol{e}_1^* \cdot \boldsymbol{E}_1)(\boldsymbol{e}_2^* \cdot \boldsymbol{E}_2) \, d\Omega$. Considering Eq. (5), one has

$$\int (\boldsymbol{e}_1^* \cdot \boldsymbol{E}_1)(\boldsymbol{e}_2^* \cdot \boldsymbol{E}_2) \, d\Omega = \sum_{k=1}^{M}(\boldsymbol{e}_1^* \cdot \boldsymbol{p}_{1,k})(\boldsymbol{e}_2^* \cdot \boldsymbol{p}_{2,k}) = [\boldsymbol{e}_1^* \boldsymbol{e}_2^*] \cdot \sum_{k=1}^{M}[\boldsymbol{p}_{1,k} \boldsymbol{p}_{2,k}]$$

$$= [\boldsymbol{e}_1^* \boldsymbol{e}_2^*](c_{h_1 h_2} [\boldsymbol{h}_1 \boldsymbol{h}_2] + c_{h_1 h_2} [\boldsymbol{h}_1 \boldsymbol{v}_2] + c_{h_1 h_2} [\boldsymbol{v}_1 \boldsymbol{h}_2] + c_{h_1 h_2} [\boldsymbol{v}_2 \boldsymbol{v}_2]) \tag{7}$$

where,

$$c_{h_1h_2} = \sum_{k=1}^{M} p_{1,k}^{H} p_{2,k}^{H}, c_{h_1v_2} = \sum_{k=1}^{M} p_{1,k}^{H} p_{2,k}^{V}, c_{v_1h_2} = \sum_{k=1}^{M} p_{1,k}^{V} p_{2,k}^{H}, c_{v_1v_2} = \sum_{k=1}^{M} p_{1,k}^{V} p_{2,k}^{V} \qquad (8)$$

In Eq. (7), $[e_1^* e_2^*]$ denotes dyadic vector generated by $e_1^*$ and $e_2^*$, and those applied for other vectors are similar. By using the compact notation $|\ )$ in the main text, one can reform Eq. (7) into $[(e_1|(e_2|]|2E)$, which is of the same form with the projection of the 2-qubit state. More generally, one can obtain the state given by Eq. (1) with $N$ correlated beams, analog to the $N$-qubit state

$$\sum_{j_1,j_2,\dots,j_N=0}^{1} q_{j_1j_2\dots j_N} |j_1 j_2 \dots j_N\rangle \qquad (9)$$

As mentioned in the main text, each of the beams is expressed by $\sum_{k=1}^{M} f_k \boldsymbol{p}_{i,k}$, such that the general expressions of the $c_{h_1h_2\dots h_N}, \dots, c_{v_1v_2\dots v_N}$ in Eq. (1) are given by,

$$c_{h_1h_2\dots h_N} = \sum_{k=1}^{M} p_{1,k}^{H} p_{2,k}^{H} \dots p_{N,k}^{H}$$

$$c_{h_1h_2\dots v_N} = \sum_{k=1}^{M} p_{1,k}^{H} p_{2,k}^{H} \dots p_{N,k}^{V}$$

$$\vdots \qquad \vdots$$

$$c_{v_1v_2\dots v_N} = \sum_{k=1}^{M} p_{1,k}^{V} p_{2,k}^{V} \dots p_{N,k}^{V} \qquad (10)$$

In the main text, we point out that $c_{h_1h_2\dots h_N}, \dots, c_{v_1v_2\dots v_N}$ in Eq. (1) have a one-to-one relationship with $q_{00\dots0}, \dots, q_{11\dots1}$ in Eq. (9). The underlying conditions for the one-to-one relationship are precisely characterized by Eqs. (10). In fact, the coefficients $c_{h_1h_2\dots h_N}, \dots, c_{v_1v_2\dots v_N}$ in Eq. (1) can be viewed as the components of a vector in the Hilbert space[40]. More generally, any other physical system that can be described by the similar mathematics would be also characterized by the Hilbert space.

### The details of the numerical simulation shown by Fig. 2

The specific models of the numerical simulation in Fig. 2 are instructed below. In Fig. 2(a), we compare the convergence of the loss function of the two networks in a binary classification task. The task we consider is to learn the labels of the eight-dimensional 0-1 vectors. The labels of the vectors are either "0" or "1". In our consideration, the labels of the vectors are generated randomly. The results shown in Fig. 2(a) are averaged over the data obtained by 12 times of training. In each training, the vector set is re-labeled randomly.

To accomplish the task, we set the COCNN by adopting one C-layer and one P-layer on three input beams, such as $\widetilde{E}_1$, $\widetilde{E}_2$ and $\widetilde{E}_3$. The state of the beams can be given by $|3E) = c_{h_1h_2h_3} |h_1\rangle|h_2\rangle|h_3\rangle +$

$c_{h_1 h_2 v_3} |\boldsymbol{h}_1\rangle|\boldsymbol{h}_2\rangle|\boldsymbol{v}_3\rangle + \cdots + c_{v_1 v_2 v_3} |\boldsymbol{v}_1\rangle|\boldsymbol{v}_2\rangle|\boldsymbol{v}_3\rangle$). The C-layer here contains two 2-beam operations shown in the dashed blue box of Fig. 1. Because the task is not so complicated, each 2-beam operation only has six trainable parameters. The 2-beam operation can be given in a matrix multiplication form,

$$[U_Z(\theta_1) \otimes U_Y(\theta_2)] \cdot U_{CX} \cdot [U_Y(\theta_3) \otimes U_Z(\theta_4)] \cdot U'_{CX} \cdot [U_Y(\theta_5) \otimes I] \cdot U_{CX} \cdot [U_Z(\theta_6) \otimes H] \quad (11)$$

whose basis is $\{|\boldsymbol{h}_1\rangle|\boldsymbol{h}_2\rangle, |\boldsymbol{h}_1\rangle|\boldsymbol{v}_2\rangle, |\boldsymbol{v}_1\rangle|\boldsymbol{h}_2\rangle, |\boldsymbol{v}_1\rangle|\boldsymbol{v}_2\rangle\}$ (or $\{|\boldsymbol{h}_2\rangle|\boldsymbol{h}_3\rangle, |\boldsymbol{h}_2\rangle|\boldsymbol{v}_3\rangle, |\boldsymbol{v}_2\rangle|\boldsymbol{h}_3\rangle, |\boldsymbol{v}_2\rangle|\boldsymbol{v}_3\rangle\}$). The notation in Eq. (11) is defined by,

$$U_Z(\theta) = \exp(-iZ\theta/2), \quad U_Y(\theta) = \exp(-iY\theta/2)$$

$$U_{CX} = \begin{pmatrix} 1 & 0 & 0 & 0 \\ 0 & 1 & 0 & 0 \\ 0 & 0 & 0 & 1 \\ 0 & 0 & 1 & 0 \end{pmatrix}, \quad U'_{CX} = \begin{pmatrix} 1 & 0 & 0 & 0 \\ 0 & 0 & 0 & 1 \\ 0 & 0 & 1 & 0 \\ 0 & 1 & 0 & 0 \end{pmatrix} \quad (12)$$

In fact, $U_Y(\theta)$ and $U_Z(\theta)$ are the Pauli-Y and Pauli-Z rotations in the group theory. $U_{CX}$ and $U'_{CX}$ are the same with the matrix form of the quantum CNOT gates. The two 2-beam operations act on the states encoded by the first pair of beams ($\widetilde{E}_1$, $\widetilde{E}_2$) and the second pair of beams ($\widetilde{E}_2$, $\widetilde{E}_3$) individually, while the parameters of the two operations are independently trained. After the C-layer, a P-layer is applied such that only one beam (such as $\widetilde{E}_2$) of the three is left, which contains the correlation information of the others. The specific formula can be given by applying Eq. (2) twice,

$$|3E\rangle\langle 3E| \rightarrow [\langle\boldsymbol{h}_1| \otimes I \otimes \langle\boldsymbol{h}_3|]|3E\rangle\langle 3E|[|\boldsymbol{h}_1\rangle \otimes I \otimes |\boldsymbol{h}_3\rangle] + [\langle\boldsymbol{h}_1| \otimes I \otimes \langle\boldsymbol{v}_3|]|3E\rangle\langle 3E|[|\boldsymbol{h}_1\rangle \otimes I \otimes |\boldsymbol{v}_3\rangle]$$
$$+ [\langle\boldsymbol{v}_1| \otimes I \otimes \langle\boldsymbol{h}_3|]|3E\rangle\langle 3E|[|\boldsymbol{v}_1\rangle \otimes I \otimes |\boldsymbol{h}_3\rangle]$$
$$+ [\langle\boldsymbol{v}_1| \otimes I \otimes \langle\boldsymbol{v}_3|]|3E\rangle\langle 3E|[|\boldsymbol{v}_1\rangle \otimes I \otimes |\boldsymbol{v}_3\rangle]$$

(13)

Notice that Eq. (13) is an analogy of looking into the second qubit of a 3-qubit system. The final output of the COCNN here is given by projecting the polarization state of beam $\widetilde{E}_2$ onto a direction and taking its modular square. The direction is parameterized by 3 real variables, corresponding to the horizontal and vertical components and their difference in phases respectively. In summary, the COCNN here has 15 trainable parameters. During the training process, the three parameters of the final projection are normalized after being updated in each iteration. In a real experimental setup, the 15 trainable parameters can be tuned by adjusting the fast-axis angles of the waveplates. Particularly, because the two 2-beam operations defined by Eq. (11) are implemented by the setup in the blue dashed box of Fig. 1, the 12 parameters of them can be tuned by the corresponding waveplates in the setup. Also, because the projection is implemented by an interferometer with an LO input whose polarization is at the direction of projection, the three parameters of projection are tuned by the corresponding waveplates for modulating the polarization of the LO input, as illustrated in the detection part of

Fig. 1.

In order to perform a fair comparison, the CNN we employ also has one layer for traditional convolution and one layer for traditional pooling, with 15 parameters in total. Because the input is an $8 \times 1$ vector, the convolution kernel is set to be a $3 \times 1$ vector, the elements of which are trainable. We apply three independent kernels in the layer for convolution, so the parameter number of the layer is 9. After the convolution, one input vector is transformed into three $6 \times 1$ vectors. Then, the data is fed into the layer for pooling. We apply max pooling here, and the strip is set to be 3. Then, three $4 \times 1$ vectors are obtained. The final output is given by the weighted sum of the three vectors modulated by the sigmoid function. In specific, the elements of each vector are firstly summed up so that three values in total are obtained. Then, each of the three numbers are multiplied by one parameter and added by one bias respectively, so three new values are obtained. In the end, the three new values are substituted into the sigmoid function, and the average of the sigmoid function outputs is used as the final output of the network.

In Fig. 2(b), we compare the convergence of the loss function of the two networks in a more complicated task, the multiclass classification task. The task we consider is to learn the labels of the sixteen-dimensional 0-1 vectors. The labels of the vectors are "0", "1", "2" or "3". In our consideration, the labels of the vectors are generated randomly. The results shown in Fig. 2(b) are averaged over the data obtained by 12 times of training trials. In each trail, the vector set is also re-labeled randomly. The fundamental setups of the COCNN and the reference CNN are similar to those applied for the cases of Fig. 2(a). The COCNN in this case also adopts one C-layer and one P-layer, acting on four input beams. The C-layer contains three 2-beam operations defined by Eq. (11). They operate on the $1^{st}$-$2^{nd}$, the $2^{nd}$-$3^{rd}$, and the $3^{rd}$-$4^{th}$ beams respectively, and all the parameters are trained independently. The P-layer contains two combiners, acting on the $1^{st}$-$2^{nd}$ and the $3^{rd}$-$4^{th}$ beams. Therefore, only two beams are left. The correlated measurement of the two beams is parameterized by 6 real variables in total. Each local projection is parameterized by 3 real variables as the above case. Adding them all, the whole COCNN here has 24 trainable parameters. During the training process, the three parameters of the final projection are also normalized after being updated in each iteration. The CNN here also has one layer for traditional convolution and one layer for traditional pooling, with 24 parameters in total. With the $16 \times 1$ input, the convolution kernel is set to be $5 \times 1$ and $6 \times 1$ vectors, the elements of which are trainable. We apply two $5 \times 1$ kernels and one $6 \times 1$ kernel, which are independently trained. After the convolution, one input vector is transformed into two $12 \times 1$ vectors and one $11 \times 1$ vector. Then, the data is fed into the layer for pooling. We apply average pooling here, and the strip is set to be 4. Then, two $9 \times 1$ vectors and one $8 \times 1$ vector are obtained. The final output is given by the weighted sum of the three vectors modulated by the sigmoid function. In specific, the elements of the three vectors are combined to a $26 \times 1$ vector, and then divided into four sets. Three of the sets have six elements, one has eight elements. Then, sum up the values of each set, and multiply the outcomes by four parameters with four biases being added respectively. Hence, four new values are obtained. In the end, the four new values are substituted into the sigmoid function, and the average of the sigmoid function outputs is used as the final output of the network.

In summary of the setup for Fig. 2(a) and 2(b), the COCNN model for the task contains one C-layer and one P-layer, with 15 and 24 parameters respectively. As the reference, the CNN model for the same task contains one layer for convolution and one layer for pooling, with the same number of parameters correspondingly as well. Therefore, a faster convergence of the loss function of the COCNN than that of the CNN is observed. As mentioned in the main text, it is comparable with the results in Ref. (39).

In Fig. 2(c), we further show the connection of the COCNN with the QCNN by simulating the phase recognition circuit in Ref. (37). The task in this case is to identify the phase of the input states, which are the ground states of the different Haldane models. The Hamiltonian of the Haldane model is given by Eq. (4). We numerically simulate the Hamiltonian when the number of sites is 12, which is sufficient to show the phase boundary according to our results in Fig. 2(c). Based on Eq. (1) and (10), the ground states can be encoded by the 12 correlated beams discussed above. Then, the beams are modulated by the COCNN analogy to the QCNN shown in the Fig. 2(b) of Ref. (37). In the main text, we mentioned that the C-layer is composed of a series of 2-beam operations. Each operation can generate any type of correlated states of the two beams, so it can realize the analogy of all the quantum gates in the QCNN. The essential gates of the QCNN in Fig. 2(b) of Ref. (37) are CZ gates, Toffoli gates, SWAP gates, and measurement-based phase-flip gates. The corresponding operations of our COCNN scheme are thoroughly discussed in S3 of the Supplementary Information which specifies the connections of two networks. After being modulated by the composite of one C-layer, one P-layer, and an additional fully connected layer, only four beams are left and measured on their $X$-basis. If the four output beams are denoted by $\widetilde{\boldsymbol{E}}_4$, $\widetilde{\boldsymbol{E}}_5$, $\widetilde{\boldsymbol{E}}_6$, and $\widetilde{\boldsymbol{E}}_7$ and the state of them is denoted by $|4\widetilde{E}\rangle\langle 4\widetilde{E}|$, the result of the measurement on the $X$ basis can be expressed by $\text{Tr}\{X_1 \otimes X_2 \otimes X_3 \otimes X_4 |4\widetilde{E}\rangle\langle 4\widetilde{E}|\}$. The correlation form of the measurement is given by

$$\sum_{e_4,e_5,e_6,e_7=p^-}^{p^+} (-1)^{C(e_4,e_5,e_6,e_7)} \left| \int (\boldsymbol{e}_4 \cdot \widetilde{\boldsymbol{E}}_4)(\boldsymbol{e}_5 \cdot \widetilde{\boldsymbol{E}}_5)(\boldsymbol{e}_6 \cdot \widetilde{\boldsymbol{E}}_6)(\boldsymbol{e}_7 \cdot \widetilde{\boldsymbol{E}}_7) \, d\Omega \right|^2 \quad (14)$$

where $\boldsymbol{e}_a$ ($a = 4,...,7$) here are restricted to be either $\boldsymbol{p}^- = (\boldsymbol{h}-\boldsymbol{v})/\sqrt{2}$ or $\boldsymbol{p}^+ = (\boldsymbol{h}+\boldsymbol{v})/\sqrt{2}$, and $C(\boldsymbol{e}_4,\boldsymbol{e}_5,\boldsymbol{e}_6,\boldsymbol{e}_7)$ returns the number of $\boldsymbol{p}^-$ among the $\boldsymbol{e}_a$s. By changing the coefficient ratios $h_1/J$ and $h_2/J$, one can calculate the different ground states and encode them by the 12 beams as mentioned above. Then, using Eq. (14) to do the measurements, a two-dimensional surface can be obtained by plotting the results. The red dots in Fig. 2(c) are plotted by marking the turning point of the surface in accordance with its second order derivative. The colors of background are determined by the ground-state energy density, whose boundaries are identified by the derivative of the energy density function. Notice that identifying the phase of the ground state usually requires to measure the string order parameter of all the particles. The motivation for proposing the QCNN is to decrease the number of the particles needed to measure. In our simulated example, the ground state of the 12-site Haldane Hamiltonian is mapped to the correlation of the 12 beams, and a corresponding phase graph can be obtained by measuring the four output beams. Therefore, the results can be viewed as an illustration of the QCNN spirit, showing a good match with the results in Ref. (37).

## The strict explanation of the simplification strategy used in our experiments

We perform the experiment to show a feasible COCNN analog to the 3-qubit QCNN. In general, an arbitrary three-qubit state can be denoted by

$$|\psi_3\rangle = q_{000}|000\rangle + q_{001}|001\rangle + q_{010}|010\rangle + q_{011}|011\rangle$$
$$+ q_{100}|100\rangle + q_{101}|101\rangle + q_{110}|110\rangle + q_{111}|111\rangle \quad (15)$$

Using the strategy in the first section of Materials and methods and the main text, we adopt three beams to encode Eq. (15). The number of modes is required to support the solvability of Eqs. (10). For example, if the mode number $M = 2$, one has

$$\widetilde{E}_r = (p_{r,1}^H \mathbf{h} + p_{r,1}^V \mathbf{v})f_1 + (p_{r,2}^H \mathbf{h} + p_{r,2}^V \mathbf{v})f_2$$
$$\widetilde{E}_s = (p_{s,1}^H \mathbf{h} + p_{s,1}^V \mathbf{v})f_1 + (p_{s,2}^H \mathbf{h} + p_{s,2}^V \mathbf{v})f_2$$
$$\widetilde{E}_t = (p_{t,1}^H \mathbf{h} + q_{t,1}^V \mathbf{v})f_1 + (q_{t,2}^H \mathbf{h} + q_{t,2}^V \mathbf{v})f_2 \quad (16)$$

Notice that, there are 12 unknown variables ($p^H$ and $p^V$) in Eq. (16). According to Eqs. (10), the conditions of the 12 variables for mimicking state (15) can be characterized by 8 equations. So, the equation set has a solution in principle. However, if more modes are introduced, the experimental realization of the modulations can be further simplified. Consider the case when $M = 4$, one has

$$E_r = (p_{r,1}^H \mathbf{h} + p_{r,1}^V \mathbf{v})f_1 + (p_{r,2}^H \mathbf{h} + p_{r,2}^V \mathbf{v})f_2 + (p_{r,3}^H \mathbf{h} + p_{r,3}^V \mathbf{v})f_3 + (p_{r,4}^H \mathbf{h} + p_{r,4}^V \mathbf{v})f_4$$
$$E_s = (p_{s,1}^H \mathbf{h} + p_{s,1}^V \mathbf{v})f_1 + (p_{s,2}^H \mathbf{h} + p_{s,2}^V \mathbf{v})f_2 + (p_{s,3}^H \mathbf{h} + p_{s,3}^V \mathbf{v})f_3 + (p_{s,4}^H \mathbf{h} + p_{s,4}^V \mathbf{v})f_4$$
$$E_t = (p_{t,1}^H \mathbf{h} + q_{t,1}^V \mathbf{v})f_1 + (q_{t,2}^H \mathbf{h} + q_{t,2}^V \mathbf{v})f_2 + (p_{t,3}^H \mathbf{h} + p_{r,3}^V \mathbf{v})f_3 + (p_{t,4}^H \mathbf{h} + p_{t,4}^V \mathbf{v})f_4 \quad (17)$$

The equation set can be expressed in a matrix form,

$$\begin{pmatrix} p_{r,1}^H p_{t,1}^H & p_{r,2}^H p_{t,2}^H & p_{r,3}^H p_{t,3}^H & p_{r,4}^H p_{t,4}^H & 0 & 0 & 0 & 0 \\ p_{r,1}^H p_{t,1}^V & p_{r,2}^H p_{t,2}^V & p_{r,3}^H p_{t,3}^V & p_{r,4}^H p_{t,4}^V & 0 & 0 & 0 & 0 \\ 0 & 0 & 0 & 0 & p_{r,1}^H p_{t,1}^H & p_{r,2}^H p_{t,2}^H & p_{r,3}^H p_{t,3}^V & p_{r,4}^H p_{t,4}^V \\ 0 & 0 & 0 & 0 & p_{r,1}^H p_{t,1}^V & p_{r,2}^H p_{t,2}^V & p_{r,3}^H p_{t,3}^V & p_{r,4}^H p_{t,4}^V \\ p_{r,1}^V p_{t,1}^H & p_{r,2}^V p_{t,2}^H & p_{r,3}^V p_{t,3}^H & p_{r,4}^V p_{t,4}^H & 0 & 0 & 0 & 0 \\ p_{r,1}^V p_{t,1}^V & p_{r,2}^V p_{t,2}^V & p_{r,3}^V p_{t,3}^V & p_{r,4}^V p_{t,4}^V & 0 & 0 & 0 & 0 \\ 0 & 0 & 0 & 0 & p_{r,1}^V p_{t,1}^H & p_{r,2}^V p_{t,2}^H & p_{r,3}^V p_{t,3}^H & p_{r,4}^V p_{t,4}^H \\ 0 & 0 & 0 & 0 & p_{r,1}^V p_{t,1}^V & p_{r,2}^V p_{t,2}^V & p_{r,3}^V p_{t,3}^V & p_{r,4}^V p_{t,4}^V \end{pmatrix} \begin{pmatrix} p_{s,1}^H \\ p_{s,2}^H \\ p_{s,3}^H \\ p_{s,4}^H \\ p_{s,1}^V \\ p_{s,2}^V \\ p_{s,3}^V \\ p_{s,4}^V \end{pmatrix} = \begin{pmatrix} q_{000} \\ q_{001} \\ q_{010} \\ q_{011} \\ q_{100} \\ q_{101} \\ q_{110} \\ q_{111} \end{pmatrix}$$

(18)

Because the equation set is under determined, we can fix several unknown variables while keeping the solvability of the equation set. In our experiment, we consider to set $p_r$s and $p_t$s as follows,

$$p_{r,1}^H = 1, p_{r,1}^V = 0, p_{r,2}^H = 1, p_{r,2}^V = 0, p_{r,3}^H = 0, p_{r,3}^V = 1, p_{r,4}^H = 0, p_{r,4}^V = 1$$
$$p_{t,1}^H = 1, p_{t,1}^V = 0, p_{t,2}^H = 0, p_{t,2}^V = 1, p_{t,3}^H = 1, p_{t,3}^V = 0, p_{t,4}^H = 0, p_{t,4}^V = 1$$

(19)

Then, the coefficient matrix in reference of $p_s$s changes to

$$M_s = \begin{pmatrix} 1 & 0 & 0 & 0 & 0 & 0 & 0 & 0 \\ 0 & 1 & 0 & 0 & 0 & 0 & 0 & 0 \\ 0 & 0 & 0 & 0 & 1 & 0 & 0 & 0 \\ 0 & 0 & 0 & 0 & 0 & 1 & 0 & 0 \\ 0 & 0 & 1 & 0 & 0 & 0 & 0 & 0 \\ 0 & 0 & 0 & 1 & 0 & 0 & 0 & 0 \\ 0 & 0 & 0 & 0 & 0 & 0 & 1 & 0 \\ 0 & 0 & 0 & 0 & 0 & 0 & 0 & 1 \end{pmatrix}$$

(20)

Notice that $M_s$ is a symmetric and sparse matrix. Given a unitary operator $U_3$ on $|\psi_3\rangle$, the matrix form of $U_3$ is denoted by $M_U$. Hence, the corresponding modulation on the beams that generates the analogy of $U_3|\psi_3\rangle$ can be deduced by following relations,

$$\text{qubit state:} \quad |\psi_3\rangle = \begin{pmatrix} q_{000} \\ q_{001} \\ q_{010} \\ q_{011} \\ q_{100} \\ q_{101} \\ q_{110} \\ q_{111} \end{pmatrix} \xrightarrow{U_3} U_3|\psi_3\rangle = M_U \begin{pmatrix} q_{000} \\ q_{001} \\ q_{010} \\ q_{011} \\ q_{100} \\ q_{101} \\ q_{110} \\ q_{111} \end{pmatrix} = \begin{pmatrix} q'_{000} \\ q'_{001} \\ q'_{010} \\ q'_{011} \\ q'_{100} \\ q'_{101} \\ q'_{110} \\ q'_{111} \end{pmatrix}$$

$$\text{beam state:} \quad M_s \begin{pmatrix} p_{s,1}^H \\ p_{s,2}^H \\ p_{s,3}^H \\ p_{s,4}^H \\ p_{s,1}^V \\ p_{s,2}^V \\ p_{s,3}^V \\ p_{s,4}^V \end{pmatrix} \xrightarrow{M_U} M_U M_s \begin{pmatrix} p_{s,1}^H \\ p_{s,2}^H \\ p_{s,3}^H \\ p_{s,4}^H \\ p_{s,1}^V \\ p_{s,2}^V \\ p_{s,3}^V \\ p_{s,4}^V \end{pmatrix} = M_s M_U' \begin{pmatrix} p_{s,1}^H \\ p_{s,2}^H \\ p_{s,3}^H \\ p_{s,4}^H \\ p_{s,1}^V \\ p_{s,2}^V \\ p_{s,3}^V \\ p_{s,4}^V \end{pmatrix} = M_s \begin{pmatrix} p'^H_{s,1} \\ p'^H_{s,2} \\ p'^H_{s,3} \\ p'^H_{s,4} \\ p'^V_{s,1} \\ p'^V_{s,2} \\ p'^V_{s,3} \\ p'^V_{s,4} \end{pmatrix}$$

(21)

where $M_U' = M_s^{-1} M_U M_s$. The meaning of Eq. (21) is that applying the operation $M_U'$ on $E_s$ is equivalent to performing $U_3$ on $|\psi_3\rangle$. In our experiment, we use Eq. (21) to map all the gates of the 3-qubit QCNN to the operations on $E_s$. The optical setup in Fig. 3 is the implementation of the operations. A step-by-step calculation of the circuit is provided in S4 of the Supplementary Information.

## Acknowledgements

National key R & D Program of China (2017YFA0303800); National Natural Science Foundation of China (91850205). National Natural Science Foundation of China (No.11904022).

## Author contributions

X. Z. conceived the idea. Y. S developed the theory and performed the simulations. Q. L. conducted the experiments with the help of L.-J. K. Y. S. and X. Z. wrote the paper. X. Z. supervised the overall project.

## Competing interests

The authors declare no competing interests.

## Data availability

The data that underlie the plots within the paper and other findings of this study are available from the corresponding authors on reasonable request.

## Code availability

The code used to generate simulated data and plots is available from the corresponding authors on reasonable request.

**Supplementary information** The online version contains supplementary material available at https:xxx.

**Correspondence and requests for materials** should be addressed to X. Z.